Optically Driven Magnetic Phase Transition of Monolayer RuCl$_3$

Yingzhen Tian[1*], Weiwei Gao[2*], Erik A. Henriksen[1], James R. Chelikowsky[2,3,4], Li Yang[1]

1. Department of Physics and Institute of Materials Science and Engineering, Washington University, St. Louis, MO 63130, USA
2. Center for Computational Materials, Institute for Computational Engineering and Sciences, The University of Texas at Austin, Austin, TX 78712 USA
3. Department of Physics, The University of Texas at Austin, Austin, TX 78712 USA
4. Department of Chemical Engineering, The University of Texas at Austin, Austin, TX 78712 USA

**ABSTRACT**: Strong light-matter interactions within nanoscale structures offer the possibility of optically controlling material properties. Motivated by the recent discovery of intrinsic long-range magnetic order in two-dimensional materials, which allows for the creation of novel magnetic devices of unprecedented small size, we predict that light can couple with magnetism and efficiently tune magnetic orders of monolayer ruthenium trichloride (RuCl$_3$). First-principles calculations show that both free carriers and optically excited electron-hole pairs can switch monolayer RuCl$_3$ from the proximate spin-liquid phase to a stable ferromagnetic phase. Specifically, a moderate electron-hole pair density (on the order of $1\times10^{13}$ cm$^{-2}$) can significantly stabilize the ferromagnetic phase by 10 meV/f.u. in comparison to the zigzag phase, so that the predicted ferromagnetism can be driven by optical pumping experiments. Analysis shows that this magnetic phase transition is driven by a combined effect of doping-induced lattice strain and itinerant ferromagnetism. According to the Ising-model calculation, we find that the Curie temperature of the ferromagnetic phase can be increased significantly by raising carrier or electron-hole pair density. This enhanced opto-magnetic effect opens new opportunities to manipulate two-dimensional magnetism through non-contact, optical approaches.

1. INTRODUCTION

Since the discovery of graphene, new classes of atomically thin materials, such as transition metal dichalcogenides (TMDs), black phosphorus, and many others, have been synthesized. Such materials display a wide range of remarkable mechanical, optical, and electronic properties, which only appear in their two-dimensional (2D) forms. Particular efforts have been invested in realizing ferromagnetism in 2D materials, such as applying an external strain [1-2], doping [3], introducing adatoms [4], or using proximity effects [5]. Recently, intrinsic ferromagnetism that persists in the 2D limit was observed in monolayer CrI$_3$ [6], monolayer and few-layer CrXTe$_3$ (X=Ge, Si) [7-8], VSe$_2$ [9], and Fe$_3$GeTe$_2$ [10] etc.[11-12] These materials offer new opportunities for studying spin physics in low dimensions, and for designing spintronic devices of unprecedentedly small size [12-13]. Following these seminal discoveries, several groups independently demonstrated the tunability of magnetic orders of 2D systems [14-15]. For example, they found that the interlayer coupling of bilayer CrI$_3$ can be reversibly tuned from the antiferromagnetic (AFM) to the ferromagnetic (FM) phase by changing the free-carrier concentration via gate voltage [14-15].

Beyond mechanical strain, chemical doping, and electrostatic gating, optically tuning material properties is highly preferable as no contacts are involved. The enhanced light-matter interactions in layered van der Waals (vdW) materials strengthen the possibility to optically control material properties.[16-17] Optical pumping induced structural phase transitions in TMDs, photoinduced insulator-to-metal phase transition in $VO_2$ thin film [18-20], optically tunable magnetism[21-23], and light-induced superconductivity in cuprates [24] are examples of light controlled properties in materials. Enhanced opto-piezoelectric effects have been predicted in monolayer group IV-VI materials as well. [25] The recent breakthrough in 2D magnetism suggests a search for optically tunable magnetism in 2D systems.

Among magnetic vdW materials, α-$RuCl_3$ is a good candidate to search for opto-magnetic effects. Similar to $CrXTe_3$ (X=Ge, Si) and $CrI_3$, α-$RuCl_3$ is a layered material with weak vdW interactions between layers. First-principles calculations show that α-$RuCl_3$ has small cleavage energy, and monolayer (and few-layer) $RuCl_3$ have recently been obtained through mechanical exfoliation in experiments [26-28]. Bulk α-$RuCl_3$ displays an in-plane "zigzag" antiferromagnetic (zigzag AFM) order [29] where the magnetic moments of $Ru^{3+}$ ions align ferromagnetically with other $Ru^{3+}$ in the same zigzag chain, and antiferromagnetically with those in neighboring zigzag chains. Recent research has shown that α-$RuCl_3$ is a possible system for studying the Kitaev model. This model is exactly solvable and hosts rich and exotic magnetic phenomena, including bond-dependent exchange coupling, rich quantum spin liquid phases and fractional excitations [30]. These interesting magnetic properties can possibly be retained in the monolayer limit, given that the interlayer exchange coupling is only a few percent of the intralayer exchange coupling according to *ab initio* calculations [31], and the anisotropic alignment of magnetic moments of Ru ions provides the necessary condition for long-range 2D magnetic order, according to the Mermin-Wagner theorem [7, 32].

Here, we show that unipolar doping and optical electron-hole (*e-h*) bipolar doping can be efficient approaches to switch the ground state of monolayer $RuCl_3$ from the proximate spin-liquid phase to the FM order. The critical carrier densities for realizing this magnetic phase transition are moderate, on the order of $10^{13}$ cm$^{-2}$ for electrons, holes, and photoexcited *e-h* pairs. As a result, 2D ferromagnetism can be turned on/off by electrostatic gating and, more interestingly, by optical pumping with a practical *e-h* density. Moreover, we reveal that the electron-doping-driven FM order is mainly from lattice distortions, while the hole- and *e-h* pair-driven FM orders are mainly from itinerant electrons. The estimated Curie temperature based on an Ising model is above the liquid-nitrogen temperature and can be further increased by higher carrier/*e-h* pair densities.

## 2. RESULTS AND DISCUSSION

### 2.1. Structural and magnetic properties

Monolayer $RuCl_3$ has a similar structure as monolayer chromium trihalides, such as the ferromagnetic $CrI_3$. As shown in Figures 1 (a) and (b), $Ru^{3+}$ ions bond with six nearest-neighbor $Cl^{1-}$ ions, forming edge-sharing $RuCl_6$ octahedra, which show trigonal distortions from the regular octahedron[26, 29]. The zigzag AFM spin configuration enlarges the hexagonal unit cell to be an orthorhombic supercell containing 16 atoms, as shown in Figure 1 (b). This supercell is used for our calculations and discussions unless otherwise stated.

Depending on the ratio of the Heisenberg exchange and Kitaev interaction, the solution of the Heisenberg-Kitaev model [33] corresponds to four phases with long-range magnetic orders: the zigzag, Neel, Stripy, and ferromagnetic phases in intrinsic monolayer $RuCl_3$. To elucidate the spin configurations of these phases, we schematically plot the arrangement of the local magnetic moments of $Ru^{3+}$ ions with colored arrows in Figure 1 (c). Using an effective Hubbard $U_{eff}$=U-J of 2 eV [31, 34], our *ab initio* density functional theory (DFT) calculations reveal that the zigzag AFM phase has the lowest energy among four magnetic phases. The competing FM phase has a slightly higher energy of ~0.05 meV per f.u. (formula unit) relative to the zigzag phase. The energies of the Neel and stripy AFM phases are notably higher (more than 4.6 meV/f.u.) than the other two. These relative energies are summarized in Figure 1 (c) and agree well with a previous work [34]. By checking the effects of Hubbard $U_{eff}$ value on our calculations, we find the energy ordering of four magnetic phases (i.e., E(Zigzag) ≈

E(FM)<E(stripy)<E(Neel)) remain the same for the range of U$_{eff}$ from 2 to 5 eV (See Figure S1 in Supporting Information).

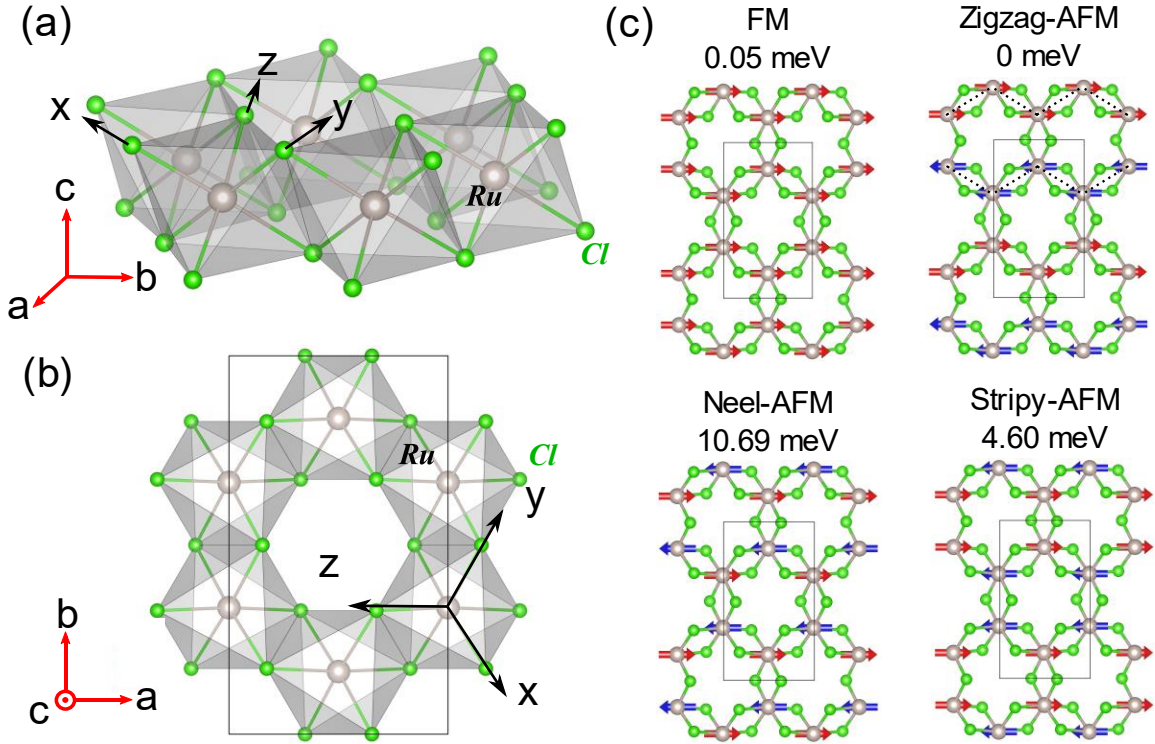

**Figure 1** (a) and (b) Top and side views of the crystal structure of monolayer RuCl$_3$. The rectangle shows the 16-atom supercell used in our calculations. (c) Schematic plots of the supercell and spin configurations of four magnetic orders, i.e., the ferromagnetic, zigzag AFM, Neel-AFM, and stripy AFM phases. The zigzag chains in zigzag-AFM phase are shown as dashed lines. The relative energies (per formula unit) to the zigzag AFM state are presented as well.

Although our calculation shows that the zigzag AFM phase has the lowest energy among four spin configurations, the small energy difference between the zigzag AFM and FM phases is only 0.05 meV/f.u., within the accuracy of the DFT+U approach. This indicates that the zigzag AFM phase and FM phase are nearly degenerate in the undoped situation. Debates on the magnetic order of monolayer RuCl$_3$ also exists in the literature. For example, Iyikanat et al. predicted that the zigzag AFM phase is the ground state [34], while Sarikurt et al. predicted that the competing FM phase is the ground state [35]. Nevertheless, both works show the energy difference between the zigzag AFM and FM phases is small (less than 1.2 meV/f.u.). The discrepancy between different calculations is due to the enhanced magnetic frustration in single-layer RuCl$_3$, as shown in a recent Raman scattering experiment[28]. In particular, as the dimension of RuCl$_3$ changes from bulk to monolayer, the experiment[28] shows the increasingly strong phonon-magnetic scattering, a hallmark of quantum spin liquids. These experimental and theoretical evidences support that the undoped monolayer RuCl$_3$ is in a proximate quantum spin-liquid phase.

### 2.2. Effects of unipolar doping

We consider the role of unipolar doping on the magnetic order of monolayer RuCl$_3$. The doped free carriers are introduced by the electrostatic doping approach, which has been widely applied to studying free-carrier doping effects on magnetism [36-38] and realized by voltage gating approaches in experiments[10, 14]. In Figure 2 (a), we present the total energies of three AFM phases (*i.e.*, zigzag, Neel, and stripy phases) relative to that of the FM phase (denoted as E(AFM)-E(FM)) under different doping densities. The Neel and stripy AFM phases always have significantly higher energies within a reasonable range of doping densities. Therefore, we focus on the competition between the FM and zigzag AFM phases.

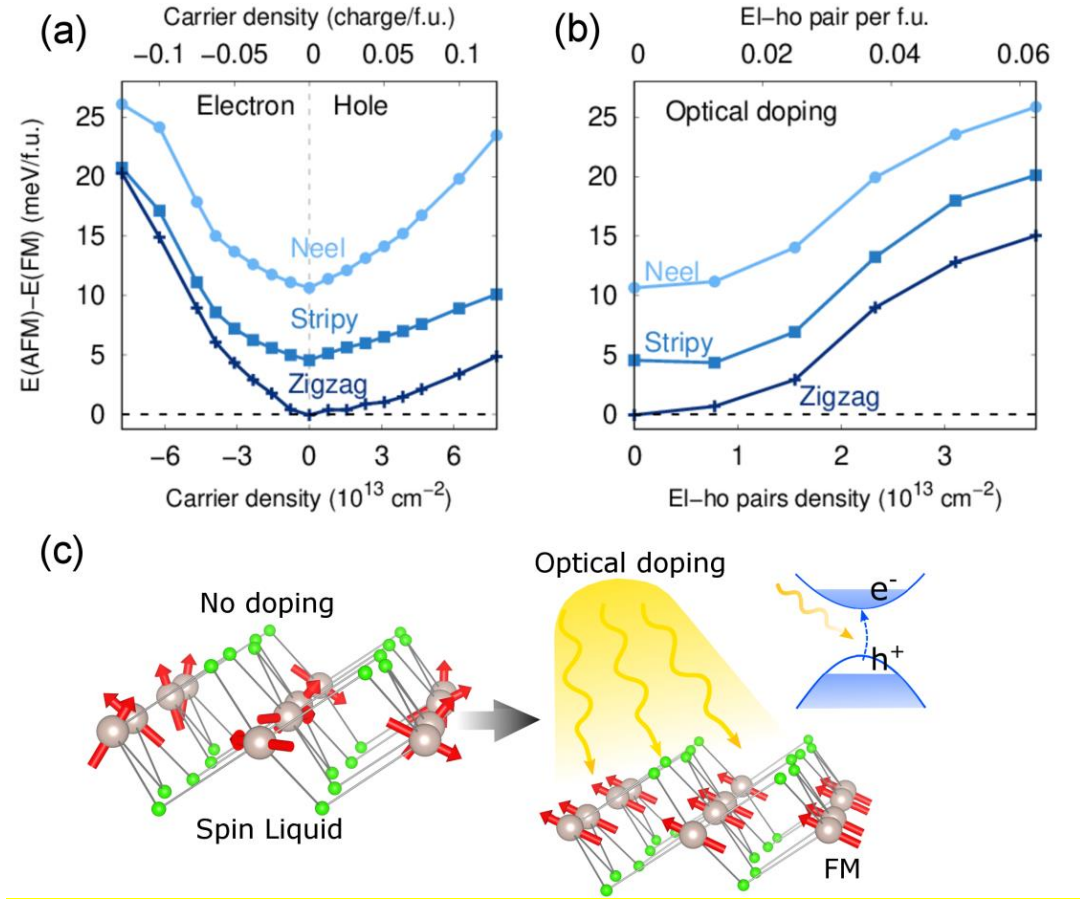

**Figure 2** (a) and (b) the relative energies to the zigzag AFM phase of monolayer RuCl3 for unipolar doping and optical *e-h* doping, respectively. (c) Schematically plot the optically tunable 2D magnetism in RuCl3.

As shown in Figure 2 (a), E(Zigzag)-E(FM) changes from a small negative value to positive as the concentration of free carriers (either electrons or holes) increases. This suggests that doped monolayer $RuCl_3$ energetically prefers the FM phase over the zigzag AFM phase. For example, to induce an E(Zigzag)-E(FM) > 5 meV/f.u., one needs to dope around $4 \times 10^{13}$ cm$^{-2}$ electrons or $8 \times 10^{13}$ cm$^{-2}$ holes. Such carrier densities are comparable with the critical density ($2.6 \times 10^{13}$ cm$^{-2}$) for the doping-induced magnetic phase transition of bilayer $CrI_3$ [14-15]. Particularly, the energy difference between the zigzag AFM and FM phases can be increased to around 20 meV/f.u. with an electron doping density of $7.7 \times 10^{13}$ cm$^{-2}$. For reference, a recently experimentally confirmed room-temperature 2D magnet $VSe_2$ displays an energy difference of 24 meV/f.u. between the FM and AFM orders by first-principles calculations[1]. Accordingly, we expect that unipolar doping can drive the ground state from the nearly degenerate magnetic orders to the FM order. Finally, we have checked the doping and screening impacts on the effective onsite Coulomb interaction U. We found that the variation of U between doping densities of 0.20 electron/supercell and 0.20 hole/supercell is about 0.2 eV. Therefore, this small variation of U will not change the above conclusion.

*2.3. Effects of optical doping*

An important character of RuCl$_3$ is that, unlike bilayer CrI$_3$ [14] and monolayer GaSe [36], in which only one type of free carriers can tune the magnetic order, monolayer RuCl$_3$ transits to the FM phase by either electron doping or hole doping. This indicates a possible optical doping effect [39-41] in which photoexcited electrons and holes are created simultaneously to drive the magnetic phase transition. To study the role of photo-excited electrons and holes, we employ a constrained DFT approximation, in which we manually change the occupation numbers of valence and conduction states to mimic the simultaneous existence of photo-excited electrons and holes. This approach has successfully been applied to examine excited-state structures of defects [42-43] and photo-induced structural changes (perovskites and monolayer SnSe) [25, 44]. The effects of optical doping are modeled with 64-atom supercells.

Figure 2 (b) illustrates the total energies of three AFM phases relative to the FM phase (denoted as E(AFM)-E(FM)) as a function of optical doping. Focusing on the competition between the FM and zigzag AFM phases, E(Zigzag)-E(FM) increases monotonically with the density of *e-h* pairs. A significant energy difference between the FM and zigzag AFM phases of approximately 10 meV/f.u. is observed under a moderate *e-h* pair density of 3.0×10$^{13}$ cm$^{-2}$. This suggests the possibility to tune magnetic properties of 2D RuCl$_3$ by non-contacting optical approaches. We plot such a photo-induced magnetic effect schematically in Figure 2 (c).

Next, we discuss the feasibility of observing photo-induced magnetic phase transition in experiments. In monolayer transition metal dichalcogenide [41, 45], optical pumping can generate *e-h* pairs with densities up to 10$^{14}$ cm$^{-2}$, which could be large enough to stabilize the FM phase. The photo-induced ferromagnetism can then be detected with the time-resolved magneto-optical Kerr effect, which measures magnetic responses in the timescale of femtosecond. [46-47] Another way to observe the photoinduced ferromagnetism in monolayer RuCl$_3$ is to apply continuous light illumination, which can ensure the photo-excited carriers to reach equilibrium. This approach has been demonstrated experimentally. By using light-emitting diode, B. Náfrádi et al. demonstrate the melting of ferromagnetism in CH$_3$NH$_3$(Mn:Pb)I$_3$[21] due to the photo-induced carriers.

Finally, the imbalance of electron and hole concentration caused by extrinsic factors, such as deep-level defects [45], are not considered in our calculations. However, as seen in Figure 2 (a), unipolar doping also strongly favors the FM order, making our prediction robust to different concentrations of electrons and holes.

*2.4. The mechanism of the doping and opto-magnetic effects*

It is important to understand the mechanism of the doping-induced effect on magnetic orders. First, we realize that the structural variation induced by doping could be an important factor to favor the FM order. Doped carriers can substantially change the lattice constants of monolayer RuCl$_3$, and strain is also known to impact the magnetic order[31, 34]. Within the range of ±0.075 electron/f.u. doping density, as shown in Figure 3 (a), the in-plane lattice constants, *a* and *b*, monotonically vary within 2%. When the type of free carriers changes from hole to electron, the in-plane strain changes from compressive to tensile. Interestingly, as shown in Figure 3 (b), if we fix the crystal structure as that of the undoped situation, electron doping does not favor the FM order. Therefore, electron doping induces a tensile strain that is crucial to favoring the FM order. On the other hand, the hole doping shrinks the in-plane lattice constants (Figure 3 (a)), and, moreover, *e-h* optical doping does not change the lattice constants significantly, as shown in Figure 3 (c). Even if we fix the crystal structure, hole doping and

*e-h* optical doping still change the relative energies between the FM and zigzag AFM orderings, as shown in Figure 3 (b) and (d), indicating that they have different mechanism to prefer the FM order.

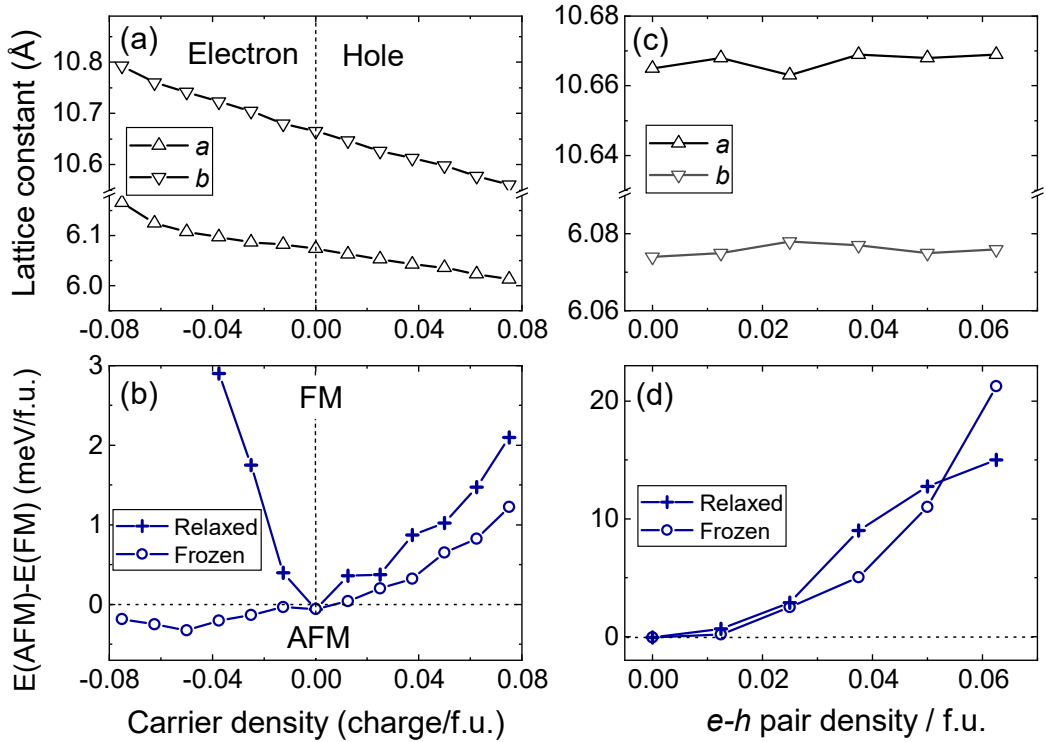

**Figure 3** (a) and (c) The lattice constants of the supercell of monolayer RuCl3 according to the unipolar and optical dopings, respectively. (b) and (d) The energy difference between the FM and zigzag AFM phases according to the unipolar and optical dopings, respectively. Results calculated with fully relaxed and frozen structures under doping are compared. Note the data points for relaxed structures (plotted with small cross) in (b) and (d) are the same as Figure 2 (a) and (b), respectively.

Many factors influence the magnetic order of solids. For doped monolayer RuCl$_3$, we identify that both itinerant electrons and localized magnetic moments contribute significantly to the doping induced ferromagnetism.

For the first mechanism, the itinerant magnetism, we calculated the density of state (DOS) for monolayer intrinsic RuCl$_3$ in a non-magnetic state. Van Hove singularities (vHSs) in the DOS appear right above and below the Fermi energy, as shown in Figure 4(a), and the vHS below the Fermi energy has a particularly higher peak than that above the Fermi energy. These vHSs have the characters of localized Ru $t_{2g}$ orbitals. According to the Stoner's theorem, if the structure is frozen as the intrinsic (undoped) case, hole doping and optical doping that also introduces holes shift the Fermi energy to vHSs with higher DOS and are more likely to induce the Stoner's instability than electron doping. This qualitatively explains why hole doping and optical doping favors the FM order even if we froze the structure at the intrinsic situation.

In addition to itinerant magnetism, the direct and indirect exchange couplings between localized moments of Ru$^{3+}$ also affect the magnetic order of doped monolayer RuCl$_3$. This is particularly important for the case of electron doping. Rau et al. [33] proposed a generic spin model to study magnetic phase diagrams for materials with honeycomb layered structures similar to the A$_2$IrO$_3$ (A=Na, Li) family iridates and RuCl$_3$. We will employ this analytical model to analyze the couplings between localized magnetic moments in carrier or optically doped monolayer RuCl3. This spin model considers the Heisenberg interaction (J), Kitaev interaction (K), and symmetric off-diagonal exchange interaction (Γ) for the honeycomb lattice.

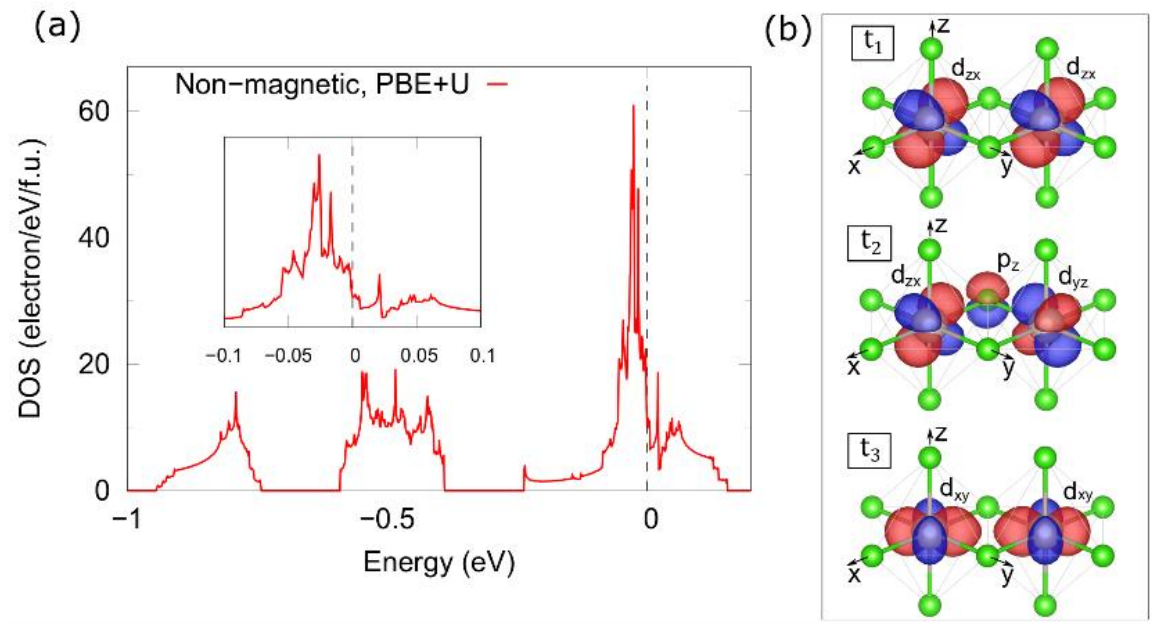

**Figure 4** (a) Density of states (DOS) of intrinsic monolayer RuCl3 in a non-magnetic state. This plot is calculated with the structure relaxed with the Zigzag AFM phase. Inset: the zoom-in plot of DOS near the Fermi energy. (b) Schematic plots of the hopping channels between t2g orbitals.

These magnetic interactions (J, K, and Γ) are determined by three hopping integrals, denoted as $t_1$, $t_2$, and $t_3$, between Ru $t_{2g}$ orbitals, which was schematically plotted in Figure 4 (b). We construct the tight-binding Hamiltonian using maximally localized Wannier functions[48] for the $t_{2g}$ orbitals of Ru atoms and obtain the hopping integrals $t_i$ (i=1,2,3) from the tight-binding Hamiltonian. Then, we calculate the exchange parameters J, K, and Γ for a few different doping conditions, as shown in Table 1. Details on calculating hopping integrals and exchange parameters are presented in the Supporting Information.

Overall, we find the Heisenberg exchange J is ferromagnetic, while the Kitaev exchange K is antiferromagnetic. The magnitude of Kitaeve interaction K is about twice the Heisenberg interaction J. The off-diagonal term Γ is ferromagnetic and smaller than K by an order of magnitude. According to the phase diagram of the generic spin model [33], the magnetic order of RuCl3 is determined by the $\phi = \arctan(\frac{J}{K})$ and $\theta = \arctan(\frac{\sqrt{J^2+K^2}}{\Gamma})$, i.e., the ratio between J, K, and Γ. The intrinsic RuCl3 has a zigzag AFM phase. As doping carriers change from holes to electrons, the parameter φ increases monotonically from 0.635π (hole doping) to 0.646π (electron doping), during which the structure is relaxed and, thus, the corresponding strain effect is included. This drives the system towards the FM phase. This qualitatively explains that electron-doping induces ferromagnetism through strain effects and variation of exchange couplings. Meanwhile, we must admit that, although the changing trend of φ qualitatively prefers the FM order, the variation value is smaller than the threshold value in Ref. [33]. One reason can be that it is hard to get close quantitative agreements between models and first-principles calculations in correlated magnetic materials. Meanwhile, the electron-induced FM order benefits from both exchange interactions and itinerant magnetism.

**Table 1. The exchange coupling constants calculated from tight-binding hopping parameters.**

| Doping density (f.u.$^{-1}$) | J (meV) | K | Γ | φ | Θ |
|---|---|---|---|---|---|
| -0.075 (electron) | -4.69 | 9.47 | -0.91 | 0.646π | 0.527π |
| 0.0 | -5.31 | 11.46 | -1.42 | 0.638π | 0.536π |
| 0.075 (hole) | -6.10 | 13.55 | -1.92 | 0.635π | 0.541π |

*2.5. Estimation of Curie temperature*

To estimate the Curie temperature of the doping-induced FM phase, we fit the total energies of these four spin configurations with a 2D Ising model, including up to third nearest-neighbor (NN) exchange coupling, and extract the exchange coupling constants. In Figures 5 (a) and (b), we present the calculated exchange coupling constants under different doping densities. Under electron doping, the magnitude of the second NN exchange coupling $J_2$ and the third NN exchange coupling $J_3$ can be as large as 8 meV, which is comparable to the NN coupling $J_1$. Interestingly, $J_1$ remains to be positive (i.e., supporting the FM order) under different doping conditions, and increases with the carrier concentration. Previous experimental works [49] also support that the coupling between NN Ru is FM.

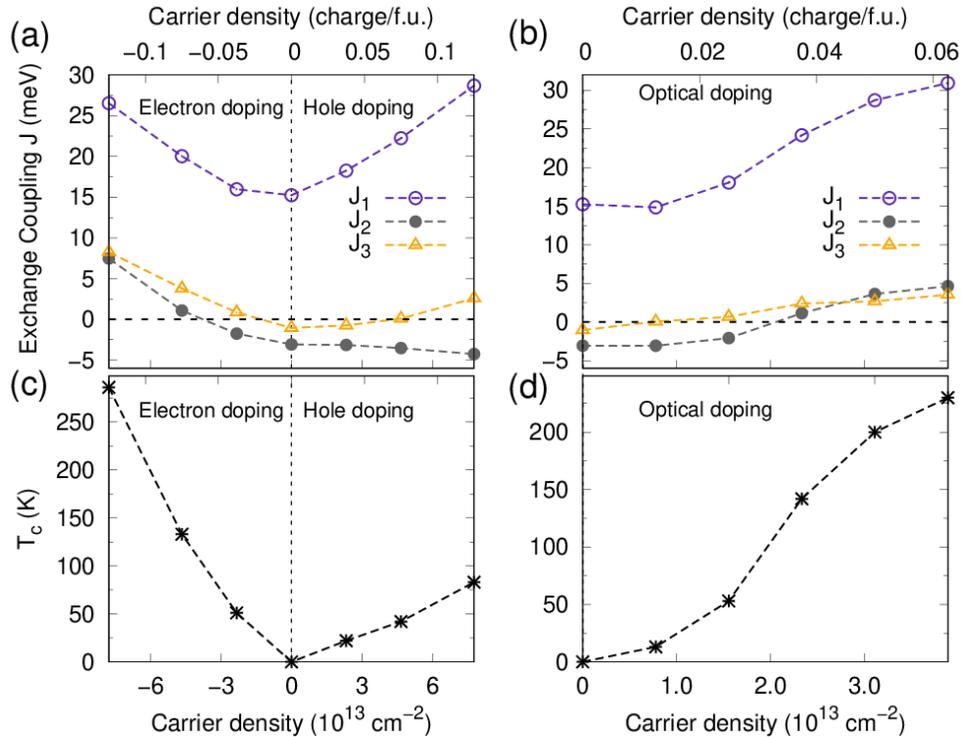

**Figure 5** (a) and (b) The exchange coupling constants ($J_1, J_2$, and $J_3$) according to the unipolar doping and optical *e-h* doping, respectively. (c) and (d) The MC simulated Curie temperature of monolayer RuCl3 under the unipolar doping and optical e-h doping, respectively.

With the exchange coupling constants $J_i$ (i=1,2,3) as inputs, we estimate the Curie temperature $T_c$ of the doping-induced FM phases using Monte Carlo (MC) simulations based on the Metropolis algorithm. Figure

5 (c) and (d) show $T_c$ for different doping densities. The estimated $T_c$ can be significantly increased under doping with the provision that the Ising model is a rough estimation, which tends to overestimate the Curie temperature.

## 3. CONCLUSION

We performed first-principles calculations to investigate the ground-state of monolayer $RuCl_3$ under electrostatic and optical dopings. Our calculations show that an intrinsic monolayer $RuCl_3$ has nearly degenerate FM and zigzag AFM orders. This results and previous work both indicate that the undoped monolayer $RuCl_3$ shows signatures of strong magnetic frustrations and quantum spin liquid. We predict that electrostatic doping with either electrons or holes and optical doping can both cause a phase transition from the spin-liquid phase to the FM order with a moderate carrier/*e-h* density, achievable with current experimental techniques. Increasing *e-h* pair density by optical doping can further enhance ferromagnetism and increases the Curie temperature significantly. The mechanisms for driving the magnetic phase transition are discussed based on changes of crystal structures and orbital components. In brief, electron doping drives the magnetic phase transition mainly by inducing a tensile strain, while hole doping and optical electron-hole doping enhances ferromagnetism by altering the occupations of Ru 4d $t_{2g}$ orbitals and itinerant magnetism. Optically driving 2D ferromagnetism offers the possibility of non-contact tunability for exploring new physics and spintronic applications.

## 4. COMPUTATIONAL DETAILS

### *4.1. DFT calculations*

We carry out pseudopotential DFT calculations with the plane-wave based Quantum Espresso code [50-51]. The Perdew-Burke-Ernzerhof (PBE) functional [52] is used in calculations. The ion core potentials (nuclei and core electrons) are described with optimized norm-conserving Vanderbilt pseudopotentials with scalar-relativistic effects, provided by the Pseudo Dojo project [53]. Semi-core 4s and 4p electrons of Ru are explicitly treated as valence electrons. Since Ru has moderate spin-orbit effects, we also did calculations for unipolar doping cases using VASP[54-56] with full relativistic effects and our main conclusions are not changed (as shown in Supporting Information).

Following previous work [31, 34], an effective Hubbard $U_{eff}$=U-J=2.0 eV is used. We notice the Hubbard $U_{eff}$ used in previous work ranges from 1.5 eV to 3.0 eV[29, 57]. An experimental study suggests that U=2.4 eV and J=0.4 eV are reasonable guesses[58]. The dependence of our results on $U_{eff}$ is studied and presented in Supporting Information.

The k-grids of 6×3×1 and 15×8×1 are used for the Brillouin-zone sampling in intrinsic and doped cases, respectively. We set the distance *d* between neighboring layers as 15 Angstrom to avoid spurious interactions. For structural optimizations, we relax both the atomic coordinates and the lattice constants so that residue forces are less than $2\times10^{-4}$ Ry/a.u. and total energies are converged within $1\times10^{-4}$ Ry. Due to the strong magnetic frustration[28], there are many competing magnetic phases of monolayer $RuCl_3$. Even though these local minima phases may have the same total magnetization, the local distributions of spin polarization can be different and result in slightly different relaxed structures and total energies. We have performed structural relaxations with different starting points to find the structure with the lowest energy.

We constructed Maximally Localized Wannier Functions (MLWF) for the Cl p orbitals and Ru $t_{2g}$ orbitals using the Wannier90 package[59-60]. The hopping integrals between Ru $t_{2g}$ orbitals is obtained from the tight-binding Hamiltonian built with MLWF.

Finally, we must point out that the first-principles simulations of correlated magnetic materials, in particular for the highly frustrated system like RuCl$_3$, are affected by the choice of functionals, pseudopotentials, effective Hubbard term U$_{eff}$, and spin-orbit coupling (SOC). In Supporting Information, we discussed their impacts and find that these factors can affect our results quantitatively. For example, the relative energies calculated by different exchange-correlation functionals can differ by 6 meV/f.u. Nevertheless, our main conclusions, i.e., the enhancement of ferromagnetism under doping conditions, still holds (see Supporting Information section 1).

*4.2. Ising model and MC simulations*

We extract the exchange coupling constants up to the third-nearest neighbor by mapping the total energies of monolayer RuCl$_3$ of four different magnetic orders to the Ising model on 2D hexagonal lattices

$$H = -\frac{1}{2}(\sum_{\langle i,j \rangle} J_1 S_i S_j + \sum_{\langle\langle i,j \rangle\rangle} J_2 S_i S_j + \sum_{\langle\langle\langle i,j \rangle\rangle\rangle} J_3 S_i S_j)$$

where $S_{1,2} = \pm 1/2$ and $J_1$, $J_2$, and $J_3$ correspond to exchange coupling constants between the NN, second NN, and third NN couplings.

With the Ising model, the exchange energies for four magnetic phases are given by:

$$E_{FM} = \frac{1}{4}(-6J_1 - 12J_2 - 6J_3)$$

$$E_{zzAFM} = \frac{1}{4}(-2J_1 + 4J_2 + 6J_3)$$

$$E_{Neel} = \frac{1}{4}(6J_1 - 12J_2 + 6J_3)$$

$$E_{stripy} = \frac{1}{4}(2J_1 + 4J_2 - 6J_3)$$

For the MC simulation, we employ a hexagonal lattice with 60×60 spin sites, which is large enough to eliminate finite-size effects. To converge the canonical ensemble average magnetization, we ran the simulation with $1.3 \times 10^5$ Monte Carlo steps for each temperature points. We provide results of MC simulations for several doping densities in Supporting Information.

## Author information

### Corresponding Author

James R. Chelikowsky: jrc@utexas.edu
Li Yang: lyang@physics.wustl.edu

### Author Contributions

Y. T. and W. G. contributed equally to this work.

## Acknowledgement

We thank Dr. Yan Lyu and Dr. Ruixiang Fei for their helpful discussion. YT and LY are supported by the National Science Foundation (NSF) CAREER Grant No. DMR-1455346 and the Air Force Office of Scientific Research (AFOSR) grant No. FA9550-17-1-0304. EAH acknowledges support under NSF DMR-1810305. Work at the University of Texas at Austin was supported from a subaward from the Center for Computational Study of Excited-State Phenomena in Energy Materials at the Lawrence Berkeley National Laboratory, which is funded by the U.S. Department of Energy, Office of Science, Basic Energy Sciences, Materials Sciences and Engineering Division under Contract No. DEAC0205CH11231, as part of the Computational Materials Sciences Program. Computational resources are provided by the Texas Advanced Computing Center (TACC).